# Stationary bound states of spin-half particles in the Reissner-Nordström gravitational field

M.V. Gorbatenko, V.P. Neznamov[1]

RFNC-VNIIEF, 37 Mira Ave., Sarov, 607188, Russia

Abstract

We prove the possibility of existence of stationary bound states of spin-half particles in the Reissner-Nordström gravitational field using a self-conjugate Hamiltonian with a flat scalar product of wave functions.

Bound states of Dirac particles with a real discrete energy spectrum can exist both for particles above the external "event horizon", and for particles under the internal "event horizon", or the Cauchy horizon.

The Hilbert condition $g_{00} > 0$ leads to a boundary condition such that components of the vector of current density of Dirac particles are zero near the "event horizons".

Based on the results of this study, we can assume that there exists a new type of charged collapsars, for which the Hawking radiation is not present.

The results of this study can lead to a revision of some concepts of the standard cosmological model related to the evolution of the universe and interaction of charged collapsars with surrounding matter.

---

[1] E-mail: neznamov@vniief.ru

# 1. Introduction

In [1] - [3], we developed an algorithm for deriving self-conjugate Dirac Hamiltonians with a flat scalar product of wave functions within the framework of pseudo-Hermitian quantum mechanics for arbitrary, including time dependent, external gravitational fields.

It follows from single-particle quantum mechanics that if the Hamiltonian is Hermitian, if there are quadratically integrable wave functions, and if appropriate boundary conditions are specified, the self-conjugate time-independent Hamiltonians should provide for the existence of stationary bound states of particles with a real energy spectrum.

In [4], [5], we for the first time obtained non-decaying bound states of spin-half particles in the Schwarzschild gravitational field using a self-conjugate Hamiltonian with a flat scalar product of wave functions for any values of the gravitational coupling constant. In order to satisfy the Hilbert condition $g_{00} > 0$, we introduced a boundary condition such that components of the vector of particle current density are zero near the "event horizon".

For the Schwarzschild metric, in terms of quantum mechanics, the "event horizon" is in fact an infinitely high potential barrier. Schwarzschild solutions in isotropic [6] and harmonic [7] coordinates possess the same property.

In this study, similarly to [4], [5], we explore the possibility of existence of stationary bound states of Dirac particles in the Reissner-Nordström gravitational field [8], [9]. As a result of the analysis, we conclude that bound states of spin-half particles with a real energy spectrum can exist both above the external "event horizon" and under the internal "event horizon", the Cauchy horizon. Results of numerical calculations of the energy spectrum and radial wave functions will be presented in next paper.

This paper has the following structure. In Sects. 2, 3, we define the self-conjugate Dirac Hamiltonian in the Reissner-Nordström field. In Sects. 4, 5, we separate variables, define equations and asymptotics for radial wave functions. In Sects. 6, 7, 8, we define the current density of Dirac particles, prove that the Hamiltonian is Hermitian and introduce boundary conditions near the "event horizons". In Sect. 9, we discuss the case of an extreme Reissner-Nordström field and the case of naked singularity. In Conclusion we summarize the results of our analysis.

## 2. Reissner-Nordström metric

The Reissner-Nordström solution is characterized by a spherically symmetric point source of gravitational field of mass $M$ and electric field with a charge $Q$. The Reissner-Nordström metric is diagonal:

$$ds^2 = -f_{R-N} dt^2 + \frac{dr^2}{f_{R-N}} + r^2 \left( d\theta^2 + \sin^2\theta d\varphi^2 \right). \qquad (1)$$

In (1), $g_{00} = -f_{R-N}$; $g^{00} = -\frac{1}{f_{R-N}}$; $f_{R-N} = \left(1 - \frac{r_0}{r} + \frac{r_Q^2}{r^2}\right)$; $r_0 = \frac{2GM}{c^2}$; $r_Q = \frac{\sqrt{G}Q}{c^2}$, $M, Q$ are mass and charge of the point source of the gravitational and electric fields.

The Hilbert condition $(-g_{00}) > 0$ leads to the necessity of considering only positive values of $f_{R-N} > 0$.

## 3. Self-conjugate Hamiltonian of spin-half particles in the Reissner-Nordström field

Below we use the system of units $\hbar = c = 1$ and notation $\tilde{\gamma}^\alpha, \gamma^\alpha$ for global and local Dirac matrices, respectively. As local matrices we use matrices in the Dirac-Pauli representation.

In the general form, the Hamiltonian for a Dirac particle of mass $m$ and charge $(-e)$ in an external gravitational and electromagnetic fields with diagonal tetrads in the Schwinger gauge can be written as [10]

$$\tilde{H} = -\frac{im}{(-g^{00})} \tilde{\gamma}^0 + \frac{i}{(-g^{00})} \tilde{\gamma}^0 \tilde{\gamma}^k \frac{\partial}{\partial x^k} - i\left(\tilde{\Phi}_0 - ieA_0\right) + \frac{i}{(-g^{00})} \tilde{\gamma}^0 \tilde{\gamma}^k \left(\tilde{\Phi}_k - ieA_k\right). \qquad (2)$$

In (2), the sign $\sim$ over some quantities means that they are calculated using tetrads in the Schwinger gauge; $\tilde{\Phi}_\mu$ and $A_\mu$ $(\mu = 0,1,2,3)$ denote bispinor connectivities and vector-potential of the external electromagnetic field, respectively; $\tilde{\gamma}^\mu$ are global Dirac matrices. The quantities $\tilde{\gamma}^\mu$ are related with the local Dirac matrices $\gamma^{\underline{\beta}}$ by the tetrads $\left(\tilde{\gamma}^\mu = \tilde{H}^\mu_{\underline{\beta}} \gamma^{\underline{\beta}}\right)$.

Non-zero tetrads in the Schwinger gauge for metric (1) are equal

$$H^0_{\underline{0}} = \frac{1}{\sqrt{f_{R-N}}}; H^1_{\underline{1}} = \sqrt{f_{R-N}}; H^2_{\underline{2}} = \frac{1}{r}; H^3_{\underline{3}} = \frac{1}{r\sin\theta}. \qquad (3)$$

In [3], we prove that a self-conjugate Hamiltonian in the $\eta$-representation with a flat scalar product of wave functions for diagonal metrics can be derived from the expression

$$H_\eta = \frac{1}{2}\left(\tilde{H}_{red} + \tilde{H}_{red}^+\right), \quad (4)$$

where $\tilde{H}_{red}$ denotes the reduced term of the Hamiltonian (2) without bispinor connectivities $\tilde{\Phi}_\mu$.

$$\tilde{H}_{red} = \tilde{H} + i\tilde{\Phi}_0 - \frac{i}{(-g^{00})}\tilde{\gamma}^0\tilde{\gamma}^k\tilde{\Phi}_k. \quad (5)$$

Given that $A_k = 0$, $A_0 = \frac{Q}{r}$ for the Reissner-Nordström solution and considering (3), (4), (5) we obtain the following expression for the Hamiltonian $H_\eta = H_\eta^+$:

$$H_\eta = im\sqrt{f_{R-N}}\gamma_{\underline{0}} - i\gamma_{\underline{0}}\gamma_{\underline{1}}\left(f_{R-N}\frac{\partial}{\partial r} + \frac{1}{r} - \frac{r_0}{2r^2}\right) - \\ -i\sqrt{f_{R-N}}\frac{1}{r}\left[\gamma_{\underline{0}}\gamma_{\underline{2}}\left(\frac{\partial}{\partial\theta} + \frac{1}{2}\text{ctg}\,\theta\right) + \gamma_{\underline{0}}\gamma_{\underline{3}}\frac{1}{\sin\theta}\frac{\partial}{\partial\varphi}\right] - \frac{eQ}{r}. \quad (6)$$

The wave function of the Dirac equation with Hamiltonian (6) $\psi_\eta$ is related to the initial function $\tilde{\psi}$ by a similarity transformation [1] - [3].

$$\psi_\eta = \eta^{-1}\tilde{\psi}. \quad (7)$$

For the Reissner-Nordström solution:

$$\eta = \left(-g^{00}\right)^{1/4} = f_{R-N}^{-1/4}. \quad (8)$$

## 4. Separation of variables

The expression in square brackets in (6) depends only on angular coordinates, and all the other summands depend only on radial coordinate. This indicates that variables in the Dirac equation with Hamiltonian (6) can be separated similarly to the case of the Schwarzschild field [4], [5] with a replacement of $f \to f_{R-N}; E \to E + \frac{eQ}{r}$.

To separate the variables, we define the bispinor $\psi_\eta(\mathbf{r},t)$ as

$$\psi_\eta(r,\theta,\varphi,t) = \begin{pmatrix} F(r)\cdot\xi(\theta) \\ -iG(r)\cdot\sigma^3\xi(\theta) \end{pmatrix} e^{im_\varphi\varphi}e^{-iEt} \quad (9)$$

and use the following equation (see, e.g., [11])

$$\left[-\sigma^2\left(\frac{\partial}{\partial\theta} + \frac{1}{2}\text{ctg}\,\theta\right) + i\sigma^1 m_\varphi\frac{1}{\sin\theta}\right]\xi(\theta) = i\kappa\xi(\theta). \quad (10)$$

In order to receive Eq. (10) we made an equivalent replacement of matrices in Hamiltonian (6):

$$\gamma_1 \to \gamma_{\underline{3}}, \quad \gamma_{\underline{3}} \to \gamma_2, \quad \gamma_2 \to \gamma_1 \tag{11}$$

In (9), (10): $\xi(\theta)$ are spherical harmonics for spin ½, $\sigma^i$ are two-dimensional Pauli matrices, $m_\varphi$ is the magnetic quantum number, and $\kappa$ is the quantum number of the Dirac equation:

$$\kappa = \pm 1, \pm 2 \ldots = \begin{cases} -(l+1), & j = l + \tfrac{1}{2} \\ l, & j = l - \tfrac{1}{2} \end{cases}. \tag{12}$$

In (12), $j, l$ are the quantum numbers of the total and orbital momentum of a Dirac particle, respectively.

$\xi(\theta)$ can be represented as in [12].

$$\xi(\theta) = \begin{pmatrix} {}_{-1/2}Y_{jm_\varphi}(\theta) \\ {}_{1/2}Y_{jm_\varphi}(\theta) \end{pmatrix} = (-1)^{m_\varphi + 1/2} \sqrt{\frac{1}{4\pi} \frac{(j - m_\varphi)!}{(j + m_\varphi)!}} \begin{pmatrix} \cos\tfrac{\theta}{2} & \sin\tfrac{\theta}{2} \\ -\sin\tfrac{\theta}{2} & \cos\tfrac{\theta}{2} \end{pmatrix} \times$$
$$\times \begin{pmatrix} \left(\kappa - m_\varphi + \tfrac{1}{2}\right) \cdot P_l^{m_\varphi - 1/2}(\theta) \\ P_l^{m_\varphi + 1/2}(\theta) \end{pmatrix}. \tag{13}$$

In (13), $P_l^{m_\varphi \pm 1/2}(\theta)$ are Legendre polynomials.

The separation of variables at $f_{R-N} > 0$ gives equations for real radial functions $F(r), G(r)$.

## 5. Equations and asymptotics for radial wave functions

A system of equations for the real radial functions $F(r), G(r)$ is written as

$$f_{R-N} \frac{dF}{dr} + \left(\frac{1 + \kappa\sqrt{f_{R-N}}}{r} - \frac{r_0}{2r^2}\right) F - \left(E + \frac{eQ}{r} + m\sqrt{f_{R-N}}\right) G = 0$$
$$f_{R-N} \frac{dG}{dr} + \left(\frac{1 - \kappa\sqrt{f_{R-N}}}{r} - \frac{r_0}{2r^2}\right) G + \left(E + \frac{eQ}{r} - m\sqrt{f_{R-N}}\right) F = 0. \tag{14}$$

In accordance with (1), we consider only positive values of $f_{R-N}$. In this case, the functions $F(r), G(r)$ are real.

We introduce dimensionless variables $\varepsilon = \dfrac{E}{m}$; $\rho = \dfrac{r}{l_c}$; $2\alpha = \dfrac{r_0}{l_c} = \dfrac{2GMm}{\hbar c}$;

$\alpha_Q = \dfrac{r_Q}{l_c} = \dfrac{\sqrt{G}Qm}{\hbar c}$; $\alpha_{em} = \dfrac{eQ}{\hbar c}$; $l_c = \dfrac{\hbar}{mc}$ is the Compton wavelength of the Dirac particle.

The quantity $f_{R-N}$ can be represented as

$$f_{R-N} = \left(1 - \frac{\rho_+}{\rho}\right)\left(1 - \frac{\rho_-}{\rho}\right), \qquad (15)$$

where

$$\rho_+ = \alpha + \sqrt{\alpha^2 - \alpha_Q^2}, \qquad (16)$$

$$\rho_- = \alpha - \sqrt{\alpha^2 - \alpha_Q^2}. \qquad (17)$$

Eqs. (14) in the dimensionless variables have the following form:

$$\left(1-\frac{\rho_+}{\rho}\right)\left(1-\frac{\rho_-}{\rho}\right)\frac{dF}{d\rho} + \left(\frac{1+\kappa\sqrt{\left(1-\frac{\rho_+}{\rho}\right)\left(1-\frac{\rho_-}{\rho}\right)}}{\rho} - \frac{\alpha}{\rho^2}\right)F -$$

$$-\left(\varepsilon + \frac{\alpha_{em}}{\rho} + \sqrt{\left(1-\frac{\rho_+}{\rho}\right)\left(1-\frac{\rho_-}{\rho}\right)}\right)G = 0$$

$$\left(1-\frac{\rho_+}{\rho}\right)\left(1-\frac{\rho_-}{\rho}\right)\frac{dG}{d\rho} + \left(\frac{1-\kappa\sqrt{\left(1-\frac{\rho_+}{\rho}\right)\left(1-\frac{\rho_-}{\rho}\right)}}{\rho} - \frac{\alpha}{\rho^2}\right)G +$$

$$+\left(\varepsilon + \frac{\alpha_{em}}{\rho} - \sqrt{\left(1-\frac{\rho_+}{\rho}\right)\left(1-\frac{\rho_-}{\rho}\right)}\right)F = 0. \qquad (18)$$

In the case $\alpha^2 > \alpha_Q^2$, $f_{R-N}$ in (15) is positive only if $\rho > \rho_+$ or $\rho < \rho_-$; the wave functions $F(\rho), G(\rho)$ are defined on the intervals $\rho \in (0, \rho_-)$ and $\rho \in (\rho_+, \infty)$. In the range of $\rho_- \leq \rho \leq \rho_+$, the wave functions are equal to zero. The quantities $\rho_+$ and $\rho_-$ are radii of the external and internal "event horizons".

If the charge is $Q = 0$ $(\alpha_Q = 0)$, then $\rho_+ = 2\alpha$; $\rho_- = 0$, and system (17) will coincide with the system of radial equations for the Schwarzschild field with one "event horizon" $r = r_0$ ($\rho = 2\alpha$).

Consider the asymptotics of the wave functions $F(\rho), G(\rho)$ as $\rho \to \infty$; $\rho \to \rho_+$ $(\rho > \rho_+)$; $\rho \to \rho_-$ $(\rho < \rho_-)$; $\rho \to 0$.

The asymptotic behavior of the wave functions for the finite motion as $\rho \to \infty$ is standard for centrally symmetric gravitational fields [12], [4], [5].

For $\rho \to \infty$ the leading terms of asymptotics equal

$$F = Ce^{-\rho\sqrt{1-\varepsilon^2}}$$
$$G = -\sqrt{\frac{1-\varepsilon}{1+\varepsilon}}F \tag{19}$$

As $\rho \to \rho_+$ $(\rho > \rho_+)$
$$F = \frac{A}{\sqrt{\rho - \rho_+}} \sin\left(M_+ \ln(\rho - \rho_+) + \varphi_+\right)$$
$$G = \frac{A}{\sqrt{\rho - \rho_+}} \cos\left(M_+ \ln(\rho - \rho_+) + \varphi_+\right) \tag{20}$$

In (20),
$$M_+ = \frac{\rho_+^2}{2\sqrt{\alpha^2 - \alpha_Q^2}}\left(\varepsilon + \frac{\alpha_{em}}{\rho_+}\right) \tag{21}$$

As $\rho \to \rho_-$ $(\rho < \rho_-)$
$$F = -\frac{B}{\sqrt{\rho_- - \rho}} \sin\left(M_- \ln(\rho_- - \rho) + \varphi_-\right)$$
$$G = \frac{B}{\sqrt{\rho_- - \rho}} \cos\left(M_- \ln(\rho_- - \rho) + \varphi_-\right) \tag{22}$$

In (22),
$$M_- = \frac{\rho_-^2}{2\sqrt{\alpha^2 - \alpha_Q^2}}\left(\varepsilon + \frac{\alpha_{em}}{\rho_-}\right). \tag{23}$$

For $\rho \to 0$ asymptotic expansions is received in [13].

In (19), (20), (22), $C, A, B, \varphi_+, \varphi_-$ are the constants of integration.

Similarly to the case of the Schwarzschild field, the oscillating functions $F$ and $G$ are ill defined at the external and internal "event horizons", but they are quadratically integrable functions at $\rho \neq \rho_+$ $(\rho > \rho_+)$ or at $\rho \neq \rho_-$ $(\rho < \rho_-)$.

## 6. Dirac particle current density

By definition, current density equals
$$j^\mu = \psi_\eta^+ \gamma_0 \gamma^\mu \psi_\eta \tag{24}$$

If we use functions (9) written as
$$\psi_\eta(\rho, \theta, \varphi) = \frac{1}{\rho} \frac{1}{\sqrt{f_{R-N}}} \begin{pmatrix} f(\rho) \cdot \xi(\theta) \\ -ig(\rho) \cdot \sigma^3 \xi(\theta) \end{pmatrix} e^{im_\varphi \varphi}, \tag{25}$$

then the current density components (24) can be represented as
$$j^0 = \psi_\eta^+ \psi_\eta = \frac{1}{\rho^2} \frac{1}{f_{R-N}}\left(f^2(\rho) + g^2(\rho)\right)\left[\xi^+(\theta)\xi(\theta)\right], \tag{26}$$

$$j^r = \psi_\eta^+ f_{R-N} \gamma_{\underline{0}} \gamma^3 \psi_\eta = -i\frac{1}{\rho^2} f(\rho) g(\rho) \left[\xi^+(\theta)(\sigma^3 \sigma^3 - \sigma^3 \sigma^3)\xi(\theta)\right] = 0, \tag{27}$$

$$j^\theta = \psi_\eta^+ \frac{f_{R-N}^{1/2}}{\rho} \gamma_{\underline{0}} \gamma^1 \psi_\eta = -\frac{2}{\rho^3 f_{R-N}^{1/2}} f(\rho) g(\rho) \left[\xi^+(\theta)\sigma^2 \xi(\theta)\right] = 0, \tag{28}$$

$$j^\varphi = \psi_\eta^+ \frac{f_{R-N}^{1/2}}{\rho \sin\theta} \gamma_{\underline{0}} \gamma^2 \psi_\eta = \frac{2}{\rho^3 \sin\theta f_{R-N}^{1/2}} f(\rho) g(\rho) \left[\xi^+(\theta)\sigma^1 \xi(\theta)\right] \neq 0. \tag{29}$$

The Eqs. (27) - (29) are received considering the equivalent replacement $\gamma$ - matrix (11).

Considering the explicit form of the angular functions (13), components $j^r(\rho)$, $j^\theta(\rho)$ is zero all over the range of variation of $\rho$, except $\rho \to 0$.

## 7. Hermiticity of the Dirac Hamiltonian in the Reissner-Nordström field

We consider Hermiticity of the Hamiltonian for two cases of the domain of $\rho$.

The first case is the domain above the "external event horizon":

$$\infty > \rho > \rho_+. \tag{30}$$

The second case is the domain under the "internal event horizon":

$$\rho_- > \rho \geq 0. \tag{31}$$

For these cases, Hamiltonian (6) is Hermitian. We can show this using the general Hermiticity condition for Dirac Hamiltonians in external gravitational fields proven in [1].

$$\oint ds_k \left(\sqrt{-g} j^k\right) + \int d^3x \sqrt{-g} \left[\psi^+ \gamma^{\underline{0}} \left(\gamma^0_{,0} + \begin{pmatrix} 0 \\ 00 \end{pmatrix}\gamma^0\right)\psi + \begin{pmatrix} k \\ k0 \end{pmatrix} j^0\right] = 0. \tag{32}$$

For a stationary centrally symmetric Reissner-Nordström field $\gamma^0_{,0} \equiv \frac{\partial \gamma^0}{\partial x^0} = 0$, Christoffel symbols $\begin{pmatrix} 0 \\ 00 \end{pmatrix}$, $\begin{pmatrix} k \\ k0 \end{pmatrix} = 0$ and condition (32) reduces to

$$4\pi\rho^2 j^r(\rho \to \infty) + 4\pi\rho^2 j^r(\rho \to \rho_+) = 0, \tag{33}$$

$$4\pi\rho^2 j^r(\rho \to \rho_-) = 0. \tag{34}$$

Eq. (27) shows that the Hermiticity condition for Hamiltonian (6) is satisfied for each of the two domains of the radial wave functions $F(\rho), G(\rho)$.

Thus, for each case, when we introduce boundary conditions, the system of equations (18) will possess a stationary real energy spectrum of bound states of spin-half particles.

## 8. Boundary conditions near the "event horizons"

Relations (27) – (29) show that the current components $j^r$ and $j^\theta$ are equal to zero, however the $\varphi$-component grows indefinitely as $\rho \to \rho_+ (\rho > \rho_+)$ and as $\rho \to \rho_- (\rho < \rho_-)$.

From this, similarly to the Schwarzschild case [4], [5], a natural boundary condition near the "event horizons" is represented by a constraint on the $\varphi$-component of Dirac current at $\rho \to \rho_+$ and at $\rho \to \rho_-$.

$$\begin{aligned} f(\rho)g(\rho)\big|_{\rho \to \rho_+} &\to 0 \quad \text{for the domain (30)}, \\ f(\rho)g(\rho)\big|_{\rho \to \rho_-} &\to 0 \quad \text{for the domain (31)}. \end{aligned} \quad (35)$$

From two possible versions of fulfilment (35) we will use the conditions

$$\begin{aligned} g(\rho)\big|_{\rho \to \rho_+} &\to 0 \quad \text{for the domain (30)}, \\ g(\rho)\big|_{\rho \to \rho_-} &\to 0 \quad \text{for the domain (31)}. \end{aligned} \quad (36)$$

Some reason for this is a smallness of function $g(\rho)$ in comparison with $f(\rho)$ in nonrelativistic approximation of the Dirac equation.

Conditions (35) determine the real energy spectrum of the system of equations (18) for the domains of the wave functions (30), (31).

## 9. Extreme Reissner-Nordström field and naked singularity

The extreme field occurs if $\alpha = \alpha_Q$ $\left(\sqrt{G}\,2M = Q\right)$. In this case, the external and internal "event horizons" coincide, their radii being equal to

$$\rho_+ = \rho_- = \alpha. \quad (37)$$

A system of equations for the radial wave functions $F(\rho)$ and $G(\rho)$ in this case is written as

$$\left(1-\frac{\alpha}{\rho}\right)^2 \frac{dF}{d\rho} + \left(\frac{1+\kappa\left|1-\frac{\alpha}{\rho}\right|}{\rho} - \frac{\alpha}{\rho^2}\right)F - \left(\varepsilon + \frac{\alpha_{em}}{\rho} + \left|1-\frac{\alpha}{\rho}\right|\right)G = 0$$

$$\left(1-\frac{\alpha}{\rho}\right)^2 \frac{dG}{d\rho} + \left(\frac{1-\kappa\left|1-\frac{\alpha}{\rho}\right|}{\rho} - \frac{\alpha}{\rho^2}\right)G + \left(\varepsilon + \frac{\alpha_{em}}{\rho} - \left|1-\frac{\alpha}{\rho}\right|\right)F = 0 \quad (38)$$

Let us take a brief look at the case of naked singularity that occurs if $\alpha_Q > \alpha$, i.e. if $Q > \sqrt{G}\,2M$. In this case, the external and internal "event horizons" disappear, the quantities $\rho_+, \rho_-$ in (16), (17) become complex, $g_{00} > 0$ for all values $\rho$. The domain of radial wave function is full space $\rho \in (0, \infty)$.

## 10. Conclusion

Based on the results of this work we come to the following conclusions:

1. For the Reissner-Nordström solution, we demonstrate the possibility of existence of stationary bound states of Dirac particles of mass $m$ and charge $(-e)$.

2. Bound states with a real discrete energy spectrum can exist both above the external "event horizon" and under the internal "event horizon", or the Cauchy horizon.

3. The external and internal "event horizons" play the role of infinitely high potential barriers, not allowing Dirac particles to cross them. The wave function of a Dirac particle in the range between the internal and external "event horizons" is zero. For the extreme Reissner-Nordström field, when $\alpha = \alpha_Q$, the role of such a barrier plays the remaining "event horizon" with radius $\rho_+ = \rho_- = \alpha$.

Based on the results of this analysis and works [4], [5], we can suppose that there exists a new type of collapsars. These collapsars are:

- inert (Dirac particles cannot pass through the "event horizons");
- have no Hawking radiation property [14] (Hawking radiation requires that there is a wave function (Dirac field operators) between the external and internal "event horizons" [15] - [22]);
- at $\alpha \geq \alpha_Q$, provide for the existence of stationary bound states of spin-half particles above the external and under the internal "event horizons".

Thus, the results of this study and works [4], [5] can be useful for the improvement of some aspects of the standard cosmological model related to the evolution of the universe and interaction of collapsars with surrounding matter.

### Acknowledgement

We thank A.L. Novoselova for the substantial technical help in the preparation of this paper.